\documentclass[12pt]{article}
\usepackage{pic03}
\usepackage{hyperref}
\usepackage{url}
\usepackage{graphicx}

\begin{document}

\title{\bf NLO SCALE DEPENDENCE OF SEMI-INCLUSIVE PROCESSES.}
\author{
A. Daleo,     C. A. Garc\'{\i}a Canal 
   \\
{\em Laboratorio de F\'{\i}sica Te\'{o}rica Departamento de
F\'{\i}sica,} \\
{\em  Facultad de Ciencias Exactas
Universidad Nacional de La Plata}\\ {\em C.C. 67 - 1900 La
Plata,  Argentina} \\
R. Sassot         \\
{\em Departamento de F\'{\i}sica, F.C.E.N,
Universidad de Buenos Aires} \\ {\em Ciudad Universitaria, Pab.1 (1428)
Buenos Aires, Argentina }}
\maketitle

\baselineskip=14.5pt
\begin{abstract}
We discuss the order-$\alpha_s^2$ gluon initiated QCD corrections 
to one particle inclusive deep inelastic processes.
We focus in the  
the NLO evolution kernels relevant for the non homogeneous QCD scale 
dependence of these cross sections and factorization.
\end{abstract}

\baselineskip=17pt

\section{Introduction}
In recent years there has been an increasing interest in 
semi-inclusive deep inelastic scattering (SIDIS), driven both by crucial 
breakthroughs in the QCD description of these processes and also 
by an incipient availability of data \cite{oad}. 

Although QCD corrections to SIDIS are well known at LO 
\cite{grau,npb}, until Ref. \cite{oad} no computations had been done up to NLO 
accuracy, nor assessments of how relevant the non homogeneous scale 
dependence might be. In LO, non homogeneous evolution effects are restricted 
to a relatively small kinematic region. This suggested to neglect 
these effects in many phenomenological analyses of polarized SIDIS, leading 
baryon production, and diffractive DIS \cite{ppp}. 

In NLO the above mentioned kinematical restrictions 
are no longer present, which in principle may lead to important corrections.
At variance with the totally inclusive case, for the 
computation of the SIDIS NLO corrections it is necessary to keep additional 
variables unintegrated. This leads to entangled singularities in more than 
one variable which requires special prescription techniques \cite{oad}.

\section{${\cal O}(\alpha_s^2)$ corrections.}

The cross section for a one-particle inclusive process in which a lepton 
scatters off a nucleon and a hadron is tagged in the final state can be 
written as \cite{grau}
\begin{eqnarray}\label{eq:hadcsec}
&&\frac{d\sigma}{dx_B\, dy\, dv_h\, dw_h}= \nonumber \\
&&\sum_{i,j=q,\bar{q},g}\int_{x_B}^{1}
	\frac{du}{u}\int_{v_h}^{1} \frac{dv_j}{v_j}\int_{0}^{1}dw\,
	f_{i/P}\bigg(\frac{x_B}{u}\bigg)\,D_{h/j}\bigg(\frac{v_h}{v_j}\bigg)
	\,\frac{d\hat{\sigma}_{ij}}{dx_B\, dy\, dv_j\, dw_j}\,
	\delta(w_h-w_j)\nonumber\\
&&	+\sum_{i}\int_{\frac{x_B}{1-(1-x_B)v_h}}^{1}
	\frac{du}{u}\,\,M_{i,h/P}\left(\frac{x_B}{u},(1-x_B)v_h\right)
	\,(1-x_B)
	\,\frac{d\hat{\sigma}_{i}}{dx_B\, dy}\,\delta(1-w_h)
	\,,
\end{eqnarray}
where in addition to the usual DIS variables $x_B$ and $y$, we introduce 
energy and angular variables $v_{h}=E_{h}/E_{0}(1-x_{B})$ and $
w_{h}=\frac{1-\cos \theta_{h}}{2}$.
$E_{h}$ and $E_{0}$ are the energies of the final state hadron and of the
incoming nucleon in the $\vec{P}+\vec{q}=0$ frame, respectively. $\theta_{h}$ 
is the angle between the momenta of the hadron and the virtual photon in the 
same frame. 

The first term in eq.(\ref{eq:hadcsec}), contain the partonic cross section
which develop forward collinear singularities ($w_h=1$) that can not be 
factorized in the usual partonic densities and fragmentation functions 
$f_{i/P}$ and $D_{h/j}$, respectively. This divergences are factorized into
fracture functions $M_{i,h/P}$ and lead to their  non homogeneus scale 
dependence.
\begin{eqnarray}\label{eq:fractev}
&&\frac{\partial \,M_{i,h/P}(\xi,\zeta,Q^2)}{\partial \log Q^2}=
	\frac{\alpha_s(Q^2)}{2\pi}
	\int_{\frac{\xi}{1-\zeta}}^{1}\frac{du}{u}\,
	\,P_{i\leftarrow j}(u)
	\,M_{j,h/P}
	\left(\frac{\xi}{u},\zeta,Q^2\right)\\\nonumber \\
	&&\,\,\,\,\,\,\,\,\,\,\,
        +\frac{\alpha_s(Q^2)}{2\pi}\,\frac{1}{\xi}
	\int_{\xi}^{\frac{\xi}{\xi+\zeta}}\frac{du}{u}
	\int_{\frac{\zeta}{\xi}}^{\frac{1-u}{u}}\frac{dv}{v}\,
	P_{ki\leftarrow j}(u,v)
	\,f_{j/P}\left(\frac{\xi}{u},
	Q^2\right)
	\,D_{h/k}\left(\frac{\zeta}{\xi\,v},Q^2\right) \nonumber
\end{eqnarray}
The first order corrections to the one-particle inclusive cross section 
can be found in \cite{grau}. At ${\cal O}(\alpha_s^2)$ the prescription of
overlapping divergences in the region $\mbox{B0}=\{u\in[x_B,x_u],\:v\in[v_h,a],\:w\in[0,1]\}$ with $x_u=x_B/(x_B+(1-x_B)v_h)$ and $w_r=(1-v)(1-u)x_B / v(u-x_B)$, can be done using the prescription
\begin{eqnarray}
&& \hspace*{-10mm}(1-w)^{-1+\epsilon_{1}}(w_r-w)^{-1+\epsilon_{2}}\stackrel{\mbox{B0}}
{\longrightarrow} \nonumber \\
&&\hspace*{10mm}\frac{\Gamma(1+\epsilon_{1})\,\Gamma(1-\epsilon_{1}-\epsilon_{2})}
{\epsilon_{1}\,(\epsilon_{1}+\epsilon_{2})\,\Gamma(1-\epsilon_{2})}\,\delta(1-w)\,
\delta(a-v)(a-z)^{\epsilon_{1}+\epsilon_{2}}
\left(a\,(1-a)\right)^{1-\epsilon_{1}-\epsilon_{2}}
\nonumber\\
&&\hspace*{10mm}
+\frac{1}{\epsilon_{1}}\,\delta(1-w)\left((a-v)^{-1+\epsilon_{1}+
\epsilon_{2}}
\right)_{+v[z,\underline{a}]}\,\left(v\,(1-a)\right)^{1-\epsilon_{1}-
\epsilon_{2}}\,w_r^{-\epsilon_{1}}\nonumber\\
&&\hspace*{10mm}\times{}_{2}F_{1}\left[\epsilon_{1},\epsilon_{1}+\epsilon_{2},1
+\epsilon_{1};\frac{1}{w_r}\right]
+\left((1-w)^{-1+\epsilon_{1}}(w_r-w)^{-1+\epsilon_{2}}\right)_
{+w[0,\underline{1}]}\,\label{eq:prescr1}
\end{eqnarray} 
and with a similar recipe for $\mbox{B1}=\{u\in[x_B,x_u],\:v\in[a,1],\:w\in[0,w_r]\}$ \cite{oad}

\section{Scale Dependence}

\begin{figure}[hbt] 
\begin{center}
\begin{minipage}{35mm}
\begin{center}
\includegraphics[height=4.0cm,width=4.2cm]
{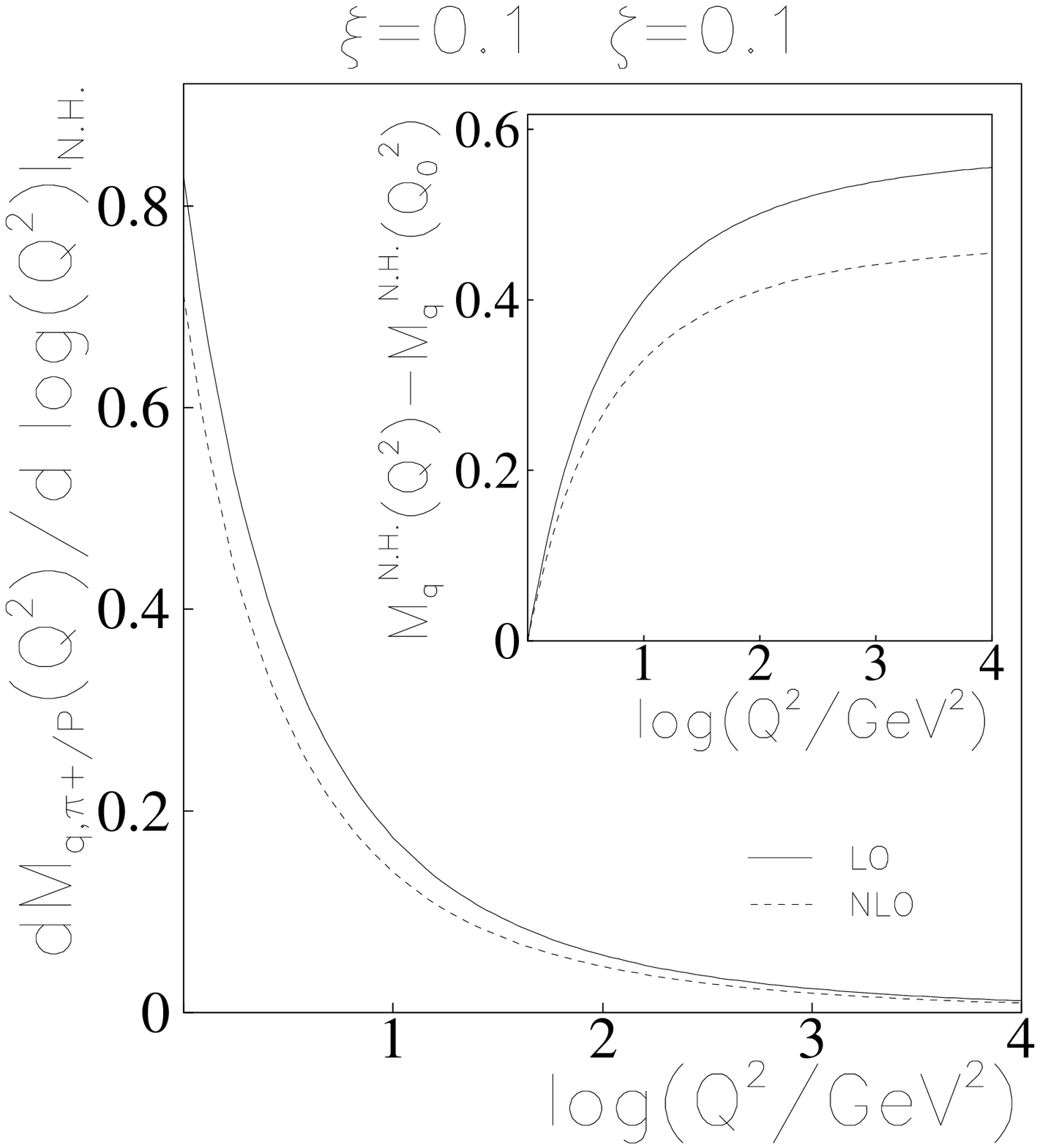}
\end{center}
\end{minipage}
\begin{minipage}{35mm}
\begin{center}
\includegraphics[height=4.0cm,width=4.2cm]
{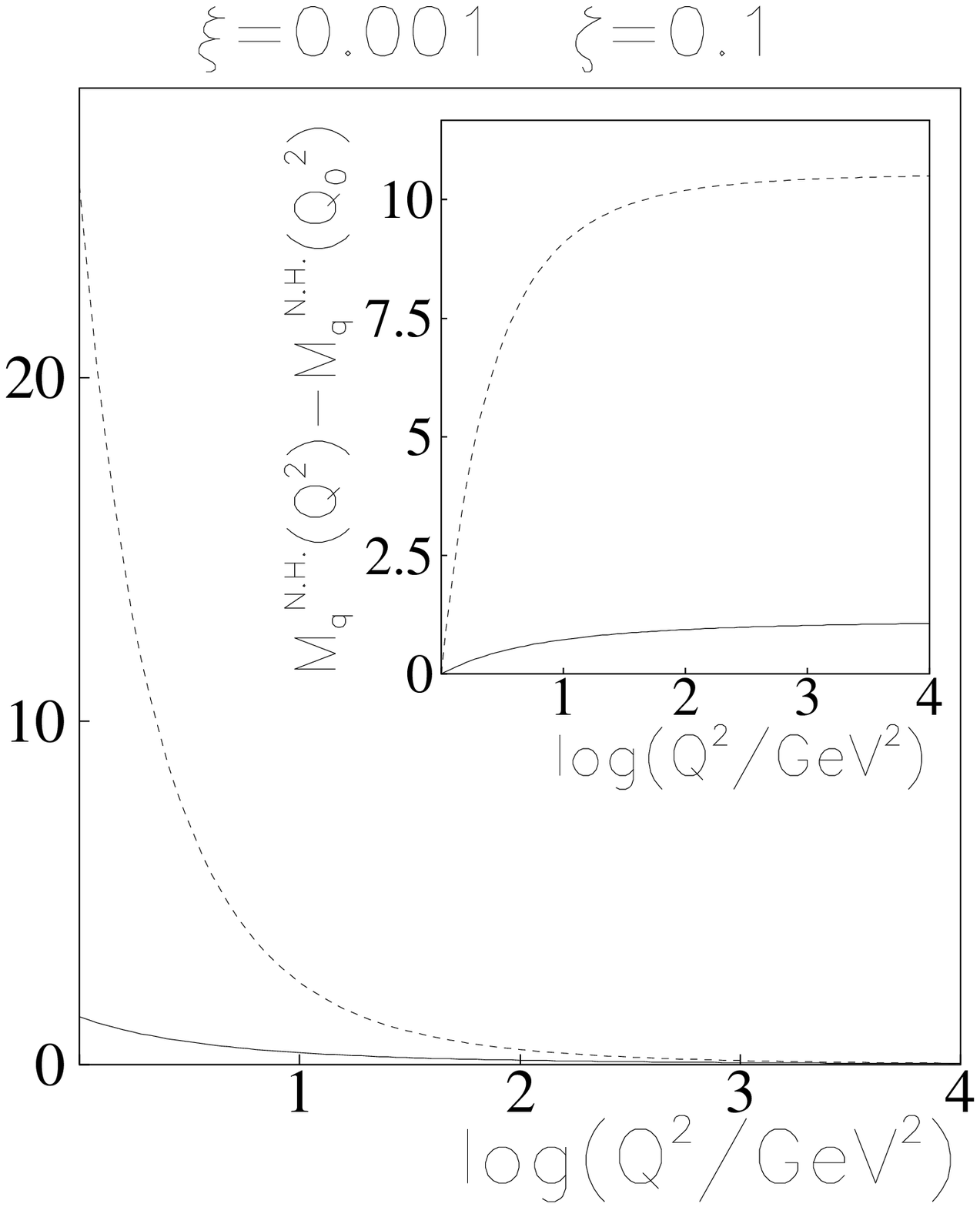}
\end{center}
\end{minipage}
\begin{minipage}{35mm}
\begin{center}
\includegraphics[height=4.0cm,width=4.2cm]
{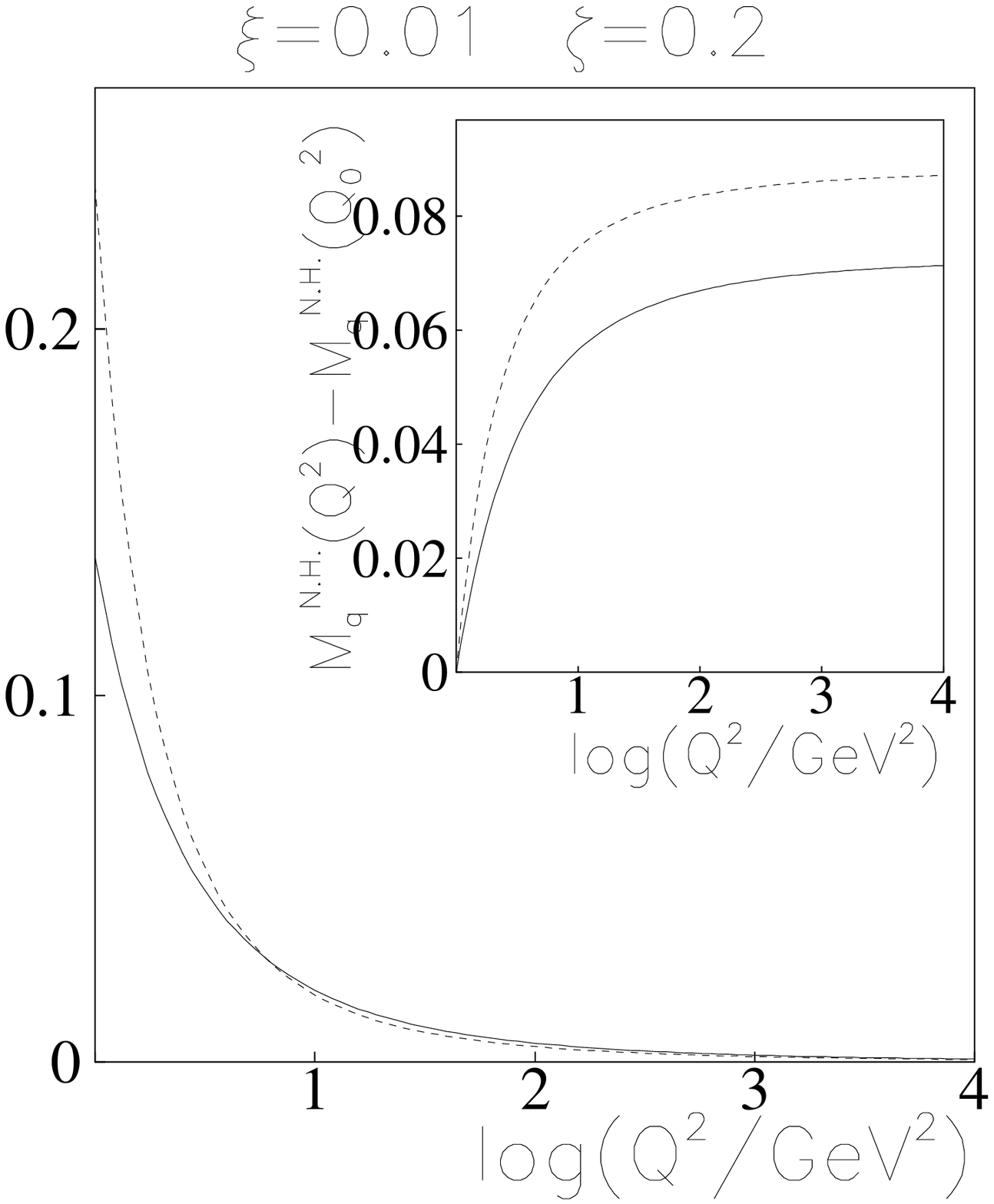}
\end{center}
\end{minipage}
\begin{minipage}{35mm}
\begin{center}
\includegraphics[height=4.0cm,width=4.2cm]
{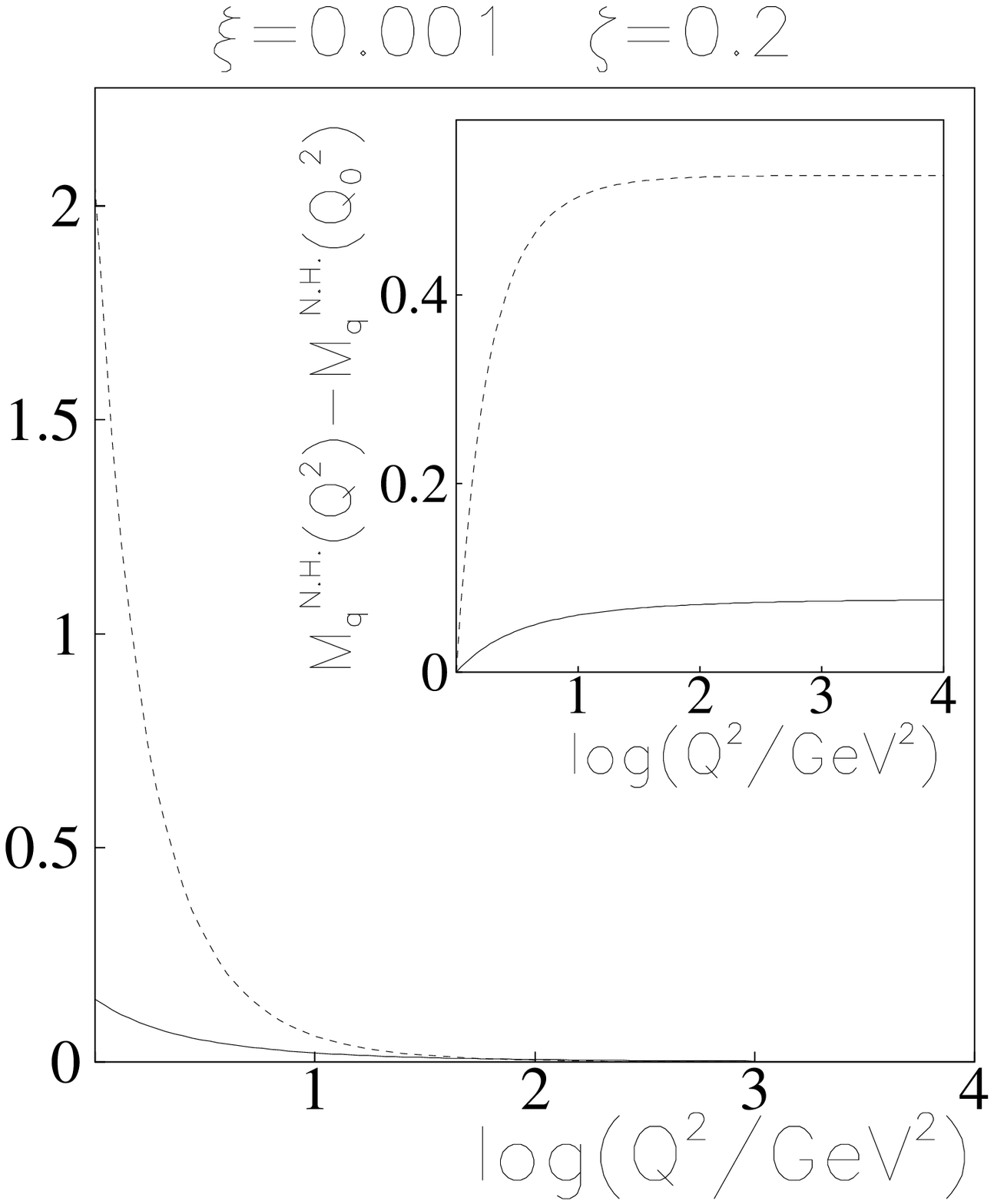}
\end{center}
\end{minipage}
\end{center}
\caption{Non homogeneous contributions to the derivative of 
$M_{q}$}
\label{fig:numcomp}
\end{figure}

Figure \ref{fig:numcomp} compares (for different
values of $\xi$ and $\zeta$) the relative
size of the LO and NLO contributions to the non homogeneous term in the 
evolution equation (\ref{eq:fractev})
computed with standard sets of parton distributions and 
fragmentation functions \cite{oad} for the case $i=q$ and $h=\pi^{+}$.
The inset plots show the integral over $Q^2$ of this contributions. 

\noindent The ${\cal O}(\alpha_s^2)$ contributions to the evolution 
equations are mild in most of the kinematical range, however they are as 
important or even larger than the
$\alpha_s$ ones for small values of $x_B$, where these last contributions are 
suppressed by the available phase space. This behavior, at variance with the
LO case, allows the non homogeneous effects to be sizable even at larger
hadron momentum fractions, thus being relevant for the scale dependence 
of a SIDIS process.

 \section{Acknowledgements}
Partially supported by CONICET, ANPCyT, UBACYT, and Fundaci\'on Antorchas.
C.A.G.C. acknowledges  the Departamento de F\'{\i}sica Teorica, Universidad de Valencia for the warm hospitality extended to him.

\end{document}